\documentclass[12pt]{article}
\input epsf
\newcommand{\R}{R}
\newcommand{\sect}[1]{\section{#1}\setcounter{equation}{0}}

\newcommand{\skyp}[1]{}

\font\mybb=msbm10 at 12pt
\def\bb#1{\hbox{\mybb#1}}
\def\Z {\bb{Z}}
\def\R {\bb{R}}

\begin{document}

\begin{titlepage}

\bigskip
\hfill\vbox{\baselineskip12pt
\hbox{NSF-KITP-03-34}
\vbox{\hbox{UCLA/03/TEP/14}}}
\bigskip\bigskip

\centerline{\Large Hidden Symmetries of the $AdS_5
\times S^5$ Superstring}

\bigskip
\bigskip
\centerline{ Iosif Bena,$^1$ Joseph Polchinski,$^2$ Radu Roiban$^3$}
\bigskip
\medskip
\centerline{$^1$\it Department of Physics}
\centerline{\it University of California, Los Angeles, CA\ \ 90095}
\centerline{ iosif@physics.ucla.edu}
\bigskip
\centerline{$^2$\it Kavli Institute for Theoretical Physics}
\centerline{\it University of California, Santa Barbara, CA\ \ 93106-4030}
\centerline{ joep@kitp.ucsb.edu}
\bigskip
\centerline{$^3$\it Department of Physics}
\centerline{\it University of California, Santa Barbara, CA\ \ 93106}
\centerline{ radu@vulcan2.physics.ucsb.edu}
\bigskip
\bigskip

\begin{abstract}
Attempts to solve  Yang-Mills theory must eventually face the problem of
analyzing the theory at intermediate values of the coupling constant.  In
this regime  neither perturbation theory nor the gravity dual are
adequate, and one must consider the full string theory in the appropriate
background.  We suggest that in some nontrivial cases the world sheet
theory may be exactly solvable. For the Green-Schwarz superstring on
$AdS_5\times S^5$ we find an infinite set of
nonlocal classically conserved charges, of the type that exist in
integrable field theories.
\end{abstract}

\end{titlepage}

\newpage
\baselineskip=18pt

\sect{Introduction}

The discovery of AdS/CFT duality~\cite{malda} was a major step towards
the longstanding goal~\cite{thooft} of recasting large-$N$ QCD as a
string theory.  The original duality was for highly supersymmetric
conformally invariant gauge theories, but these can be deformed in
various ways to produce string duals to confining gauge theories with
less or no supersymmetry.  Thus far the duality provides solutions only
for gauge theories where the 't Hooft coupling is strong at all scales,
because it is only for these that the world-sheet of the dual string is
weakly coupled.  However, it implies an existence proof (in the
physicists' sense of the term) for a string dual to QCD, by continuous
deformation to weak coupling in the UV and so to a strongly coupled
world-sheet theory.  Thus the program of solving large-$N$ QCD is reduced
to two steps:
\begin{enumerate}
\item Identify the strongly coupled world-sheet field theory of the QCD
string.
\item Solve it.
\end{enumerate} The hope is that the reduction from a strongly coupled
field theory in $3+1$ dimensions to one in $1+1$ dimensions will allow
the various special techniques of $(1+1)$-dimensional field theory to  be
brought to bear.

In this paper we will look ahead to the second step, and to report at
least one modest positive result.  Our focus here is the conformally
invariant
${\cal N}=4$ theory, where the world-sheet theory is known due to the
high symmetry of the problem, but where existing methods of calculation
have limited range.  That is, at large 't Hooft coupling we can
calculate using perturbation theory on the string world-sheet.  At small
't Hooft coupling we can calculate in the weakly coupled gauge
theory.\footnote{It is an interesting exercise to ask what this implies
about the strongly coupled world-sheet in this limit.  This will be
discussed in a separate paper~\cite{jopart}.} However, for 't Hooft
coupling of order one there are no quantitative methods, since in either
dual form the coupling is of order one. Understanding this theory should
be strictly easier than solving the  QCD string, as in the latter case
as well neither the gauge nor the string description is weakly coupled,
and there is much less symmetry.

For string theories without Ramond-Ramond (RR) backgrounds, there are
familiar strong-coupling methods based primarily on holomorphic
currents.  The Wess-Zumino-Witten model is the archetype of
this~\cite{wzw}: the spacetime symmetry is elevated to an affine Lie
algebra.  However, with RR backgrounds the bosonic Wess-Zumino term is
absent and there is no affine Lie algebra.  Also, the RR backgrounds
make it impossible to use the standard Ramond-Neveu-Schwarz conformal
field theory, within which holomorphic currents can be shown to exist
under broad conditions.

However, the Green-Schwarz superstring on $AdS_5 \times S^5$ is similar
to field theories for which other forms of higher symmetry algebra are
known to exist.  That is, the field space can be regarded as a
coset~\cite{metsey,stanford,suny},
\begin{equation}
\frac{PSU(2,2|4)}{SO(4,1)\times SO(5)}\ .
\label{supercoset}
\end{equation} For certain coset theories there are two kinds of
infinite symmetry algebras.  One is based on nonlocal
currents~\cite{LP,BIZZ,CZ} which give rise to charges satisfying
a Yangian algebra~\cite{mackay};\footnote{Transformations satisfying a
familiar affine Lie algebra can also be constructed from the nonlocal
currents~\cite{ala}.  These are invariances of the classical equations of
motion but not of the Poisson brackets (they are not symplectically
generated), and so unfortunately they probably have no quantum analogs.} for
reviews see refs.~\cite{brev,AAR}.
   The other is based on local right- or
left-moving currents~\cite{local} which satisfy a $W$-algebra; for a nice
recent discussion see ref.~\cite{EHMM}.

Of course the Green-Schwarz superstring on $AdS_5 \times S^5$ is not
precisely a coset sigma model, because of the fermionic Wess-Zumino term
and the $\kappa$ symmetry.  However, we will show that it possesses an
infinite symmetry algebra of the nonlocal form.  These nonlocal
charges are conserved in any $\kappa$ gauge.  Our considerations are
purely classical, but in related models these charges have been
argued to survive quantization~\cite{luscher}, with modified
algebras.

Our results extend immediately to the superstring on $AdS_3 \times S^3
\times T^4$ with RR flux, for which the action has the same
structure~\cite{ads3rr}.  Of course the theory with NS-NS flux is $S$-dual
to this, but in studying the world-sheet theory we are implicitly
expanding in $g_{\rm string}$ so this duality is not visible.  In fact
the NS-NS background is a case where the string side is exactly solvable,
and so an example of what we might hope to do for QCD.  Some of methods
used in this case, such as those of ref.~\cite{oogmal}, may be applicable
to gauge theories, but in some ways this example is rather different.  The
spacetime CFT is less well understood and has no adjustable coupling
constant --- thus there is no limit in which it has a classical
Lagrangian description.

In \S2 we briefly review the construction of the nonlocal charges for
bosonic nonlinear sigma models.  This is based on the identification of a
one-parameter family of flat connections, constructed from the symmetry
currents and their duals.  In \S3 we extend this to the IIB
Green-Schwarz superstring on $AdS_5 \times S^5$, and show that a
one-parameter family of flat connections exists.  In \S4 we search for
local chiral charges.  We find no charges of higher spin, but we find
that in conformal gauge the world-sheet CFT separates (at the classical
level) into two factors, one associated with $AdS_5$ and one associated
with $S^5$.  In \S5 we discuss further directions.

Beyond the application to QCD, the possibility of finding an exact
solution on one side of the gauge/gravity duality is an exciting
prospect.  Thus far these higher symmetries and related methods have
appeared in gauge/gravity in certain special contexts.  On the string
side, Maldacena and Maoz~\cite{mm} have pointed out that one can engineer
nonlinear plane wave solutions to produce an integrable world-sheet
theory (see also refs.~\cite{mmfollow}); Mandal, Suryanarayana, and
Wadia~\cite{msw} have  noted that the bosonic part of the $AdS^5 \times S_5$
string theory is classically integrable; and, Bershadsky, Zhukov, and
Vaintrob~\cite{BZW} have discussed the
$W$ symmetry of the pure supergroup sigma model. On the gauge side,
Minahan and Zarembo~\cite{mz} have shown that the calculation of
one-loop anomalous dimensions for general scalar operators can be recast in
terms of an integrable field theory (see further work~\cite{mzfollow}).
Belitsky, Gorsky and Korchemsky~\cite{BEGK} have related the computation of
anomalous dimensions certain higher spin operators to the $SL(2,\R)$ spin
chain. Also, Lipatov~\cite{lipatov} has argued that in certain QCD processes
the summation of Feynman graphs leads to an integrable model.

\sect{Review: Nonlocal charges in bosonic models}

\subsection{Principal chiral models}

Consider first the nonlinear sigma model where the field
$g(x)$ takes values in the group $G$, and the Lagrangian is
$L \propto {\rm Tr}(\partial_i g^{-1} \partial^i g)$.  The global symmetry is
$G \times G$, left and right multipication.  We will focus on the conserved
current corresponding to left multiplication,
\begin{equation} j_i = -(\partial_i g) g^{-1}\ .
\end{equation} Note that the current takes values in the Lie algebra
$\cal G$. Writing the current as a one-form, so that $d{*j} = 0$, one
sees that
\begin{equation} dj + j \wedge j = 0\ .
\end{equation} Thus the current can be regarded as a flat gauge
connection in $\cal  G$. Moreover, by taking general linear combinations
\begin{equation} a = \alpha j + \beta {*j}\ ,
\end{equation} one finds that
\begin{equation} d a + a \wedge a = (\alpha^2 - \alpha -
\beta^2) j \wedge j\ .
\label{curv}
\end{equation} We have used the identities $** = +1$, and
$*k \wedge l + k \wedge *l = 0$ for general one-forms $k, l$.  Thus
there are two one-parameter families of flat connections,
\begin{equation} a^{\lambda\pm}:\quad  \alpha =
\frac{1}{2} ( 1 \pm \cosh \lambda )\ ,\quad
\beta = \frac{1}{2} \sinh\lambda \ , \quad -\infty < \lambda < \infty\ .
\end{equation}
Equivalently, this can be written as
\begin{equation}
\alpha =
\frac{y^2}{2y - 1} \ ,\quad
\beta = \frac{y^2 - y}{2y - 1} \ , \quad -\infty < y < \infty\ .
\end{equation}

Given any flat connection, the equation
\begin{equation} d U = - a U
\end{equation} is integrable: action on both sides with
$d$ gives $0=0$.  On a simply connected space, given an initial value
$U(x_0,x_0) = 1$, this defines a group element
$U(x,x_0)$.  This is just the Wilson line, defining parallel transport
with the connection $a$,
\begin{equation} U(x,x_0) = P \exp \biggl(-\int_C a
\biggr)\ ,
\end{equation} where $C$ is any contour running from
$x_0$ to $x$, and $P$ denotes path ordering of the Lie algebra
generators.  The flatness of the connection implies that this in
invariant under the continuous deformations of $C$.

This immediately allows the construction of an infinite number of
conserved charges~\cite{LP,BIZZ}, by taking the unbounded spatial
Wilson line at fixed time,
\begin{equation} Q^{\lambda\pm}(t) = U^{\lambda\pm}(\infty,t; -\infty,
t)\ .
\end{equation}
This takes values in $G$.  The contour $C$ here is
the $t= {\rm constant}$ spatial slice, so conservation of
$Q^{\lambda\pm}$ is simply the statement that this is invariant under
continuous shift of the contour forward in time.  Of course this moves
the endpoints, which is not in general an invariance of the Wilson line,
so an appropriate falloff of the fields at infinity is assumed.
Explicitly,
\begin{eqnarray}
\partial_t U(y,t;z, t) &=&- \int_{y}^{z} dx\, U(y,t;x,t)\, {\dot
a}_1(x,t)\, U(x,t;z, t) \nonumber\\ &=& - \int_{x_1}^{x_2} dx\, U(y,t;x,
t)[a_0' - a_0 a_1 + a_1 a_0]_{(x,t)}\, U(x,t;z, t) \nonumber\\ &=&
a_0(y,t) U(y,t;z,t) - U(y,t;z,t) a_0(z,t)\ ,
\end{eqnarray} so we need $a_0(\pm \infty,t)$ to go to zero in order for
the charges to be conserved.  For closed string theory,
where the spatial direction is periodic, one takes the trace to form the
Wilson loop; for the supercoset case of the next section one would take
the supertrace.

These charges can also be presented in other forms, for example by
Taylor expanding in
$\lambda$,
\begin{equation} Q^{\lambda-} = 1 + \sum_{n=1}^{\infty} \lambda^n
Q_n \ .
\end{equation} For $a^{\lambda -}$, which vanishes at
$\lambda = 0$, we have
\begin{equation} a^{\lambda -} =  \frac{1}{2} \lambda {*j} -
\frac{1}{4}\lambda^2 j + O(\lambda^3)\ .
\end{equation} Then
\begin{equation}
2 Q_1 = \int_{-\infty}^\infty dx\, j_0(x)
\end{equation} is just the global left-multiplication charge.  Note that
we could similarly have started the whole construction with the
right-multiplication current, and gotten a second set of charges.  The
next charge is bilocal,
\begin{equation} Q_2 = -\frac{1}{4}\int_{-\infty}^\infty dx\, j_1(x)
+
\frac{1}{2}\int_{-\infty}^\infty dx \int_{-\infty}^x dx'\, j_0(x)
j_0(x')\ .
\end{equation} and so on.  Through its Poisson brackets, $Q_2$ generates
all of the higher $Q_n$, so for many purposes one can focus on this
simplest nonlocal (more precisely, multilocal) charge.

\subsection{Coset models}

Now let us consider $G/H$ coset models, where we identify
$g(x) \cong g(x) h(x)$.  Left multiplication by $G$ is still a global
symmetry.  To construct the action, define
\begin{equation} J = g^{-1} j g = - g^{-1} \partial g\ .
\end{equation} This is invariant under left multiplication. Further
separate $J$ according to the decomposition of the  Lie algebra, ${\cal
G} = {\cal H}
\oplus {\cal K}$,
\begin{equation} J = H + K\ .
\end{equation} Then $H$ transforms as a connection under
${\cal H}$-gauge transformations, whereas $K$ transforms covariantly.
It follows that
\begin{equation} k = g K g^{-1}
\end{equation} is ${\cal H}$-gauge invariant.  The Lagrangian is then $L
\propto {\rm Tr}(k_i k^i) = {\rm Tr}(K_i K^i)$.

We will use capital letters $X$ to denote currents that are conjugated
by right multiplication, generally corresponding to some decomposition
under representations of ${\cal H}$.  Then $x = g X g^{-1}$ is
conjugated by left multiplication.   We will focus on the ${\cal
H}$-gauge invariants, which are the $x$ other than $h$. Notice however
that the
$x$ do not have simple decompositions under the Lie algebra; to use such
decompositions we must refer back to the $X$.  Note also that
\begin{equation} dx = g (dX) g^{-1} - j \wedge x - x
\wedge j \ .
\end{equation}

The construction of the flat connections can be extended provided the
coset is a symmetric space~\cite{EF,AAR}. That is, in addition to
$[{\cal H},{\cal H}]
\subseteq {\cal H}$ and $[{\cal H},{\cal K}] \subseteq {\cal K}$ which
follow from the subgroup structure, we must have $[{\cal K},{\cal
K}]\subseteq {\cal H}$ as well. To see this, note that $dJ = J \wedge
J$, and decompose both sides under $G = {\cal H} \oplus {\cal K}$:
\begin{eqnarray}
dH &=& H \wedge H + K \wedge K\ ,\nonumber\\
dK &=& H
\wedge K + K \wedge H\ .
\end{eqnarray}
If the coset were not a symmetric space, then $K \wedge
K$ would be a sum of two pieces, one of which is in ${\cal H}$ and the
other in
${\cal K}$.\footnote {When ${\cal K}$ is a subalgebra,  $K
\wedge K$ contributes only to
$dK$, and it is again possible to construct flat connections.}
Transforming to the
$x$ forms, we have
\begin{eqnarray} dh &=& -k \wedge h - h \wedge k\ ,\nonumber\\ dk &=& -2
k \wedge k\ .
\end{eqnarray} The gauge invariant $k$ is also the Noether current for
the global symmetry, $d {*k} = 0$.  The current $2k$ is both flat and
conserved, and so can be used to construct two families  of flat
connections precisely as above.

The construction of the flat connections and nonlocal charges can be
extended to certain sigma models with fermions, including world-sheet
supersymmetric sigma models~\cite{CZ,AAR}.  The strategy is slightly
different: one separates the global symmetry current into its bosonic
and fermion parts, which are not separately conserved, and takes a
general linear combination of these currents and their duals.  Under
appropriate conditions there exist infinite families of flat
connections.  We will not review this here, but the construction in the
next section has a similar structure.

\sect{The Green-Schwarz superstring on $AdS_5 \times S^5$}

In order to make the construction as transparent as possible, we will
in the present subsection use a condensed notation which is parallel to
that of the earlier discussion. The Green-Schwarz superstring in
$AdS_5 \times S^5$ can be regarded as  a nonlinear sigma model where the
field takes values in the coset superspace~\cite{metsey,stanford,suny}
\begin{equation}
\frac{PSU(2,2|4)}{SO(4,1)\times SO(5)}\ . \label{supco}
\end{equation} The bosonic part of this space,
\begin{equation}
\frac{SO(4,2)}{SO(4,1)} \times \frac{SO(6)}{SO(5)} = AdS_5 \times S^5\ ,
\end{equation} is a symmetric space and so the above construction gives
an infinite symmetry algebra.  This has recently been remarked in
ref.~\cite{msw}, in the context of finding extended string solutions in
$AdS_5
\times S^5$.  The full space~(\ref{supco}) is not a symmetric
space, as  the denominator group is too small.  The theory also differs
from the  simple nonlinear sigma model by the presence of a Wess-Zumino
term, and by the
$\kappa$ gauge symmetry.  Thus the construction of the flat connection
is slightly more involved.

The Lie algebra of $PSU(2,2|4)$ can be
decomposed~\cite{metsey,stanford,suny}
\begin{equation} {\cal G} = {\cal H} + {\cal P} + {\cal Q}_1 + {\cal
Q}_2\ ,
\end{equation} where ${\cal H}$ is the denominator algebra, ${\cal P}$
contains the remaining bosonic generators, and ${\cal Q}_1$ and ${\cal
Q}_2$ are two copies of the $({\bf 4},{\bf 4})$ representation of
${\cal H}$.  The algebra respects a ${\Z}_4$ grading, under which the
charges are
\begin{equation} {\cal H} : 0\ ,\quad {{\cal Q}_1} : 1\ ,\quad {\cal P}
: 2\ ,\quad {{\cal Q}_2} : 3\ .
\end{equation} Form the current
\begin{equation} J = - g^{-1} \partial g = H + P + Q_1 + Q_2\ .
\end{equation} The grading of the Lie algebra implies that the curl $dJ
= J \wedge J$ decomposes as
\begin{eqnarray}
d H &=& H \wedge H + P \wedge P + Q_1
\wedge  Q_2 + Q_2 \wedge  Q_1\ ,\nonumber\\
d P &=& H
\wedge P + P \wedge H + Q_1  \wedge Q_1 + Q_2 \wedge Q_2\ ,\nonumber\\
  d
Q_1 &=& H \wedge  Q_1 + Q_1 \wedge H +P
\wedge  Q_2 + Q_2 \wedge P\ ,\nonumber\\
  d Q_1 &=& H
\wedge  Q_2 + Q_2 \wedge H +P \wedge  Q_1 + Q_1 \wedge P\ .
\end{eqnarray}
It is useful to define also $Q = Q_1 + Q_2$ and $Q' = Q_1 -
Q_2$, in terms of which
\begin{eqnarray}
d H &=& H \wedge H + P \wedge P + {\textstyle
\frac{1}{2}}(Q \wedge Q - Q' \wedge Q')
\ ,\nonumber\\
d P &=& H \wedge P + P \wedge H + {\textstyle
\frac{1}{2}}(Q \wedge Q + Q' \wedge Q')\ ,\nonumber\\
d Q &=& H \wedge Q
+ Q \wedge H + P \wedge Q + Q \wedge P\ ,\nonumber\\
d Q' &=& H \wedge Q' + Q'
\wedge H - P \wedge Q' - Q' \wedge P\ .
\end{eqnarray}
The curls of the lower-case forms are then
\begin{eqnarray}
d h &=& -h \wedge h + p \wedge p - h
\wedge p - p \wedge h - h \wedge q - q \wedge h + {\textstyle
\frac{1}{2}}(q \wedge q - q' \wedge q')\ ,\nonumber\\
d p &=& - 2p
\wedge p - p \wedge q - q
\wedge p + {\textstyle
\frac{1}{2}}(q \wedge q + q'
\wedge q')\ ,\nonumber\\
d q &=& -2q \wedge q\ ,\nonumber\\ d q' &=& -2p
\wedge q' -2q' \wedge p -q
\wedge q' - q' \wedge q \ .
\label{curls}
\end{eqnarray}

In the notation of ref.~\cite{metsey},
\begin{equation} H = \frac{1}{2}(L^{ab}J_{ab} + L^{a'b'}J_{a'b'})\ ,
\quad P =L^{a}P_{a} + L^{a'}P_{a}\ , \quad Q_I = L^{\alpha\alpha'I}
Q_{\alpha\alpha'I}\ .
\end{equation} Translating eqs.~3.18, 3.19 and~3.20 of that paper, the
equations of motion are\footnote{These equations, as well as eqs.(\ref{curls})
can be obtained from ref.\cite{suny} with the identifications
\begin{eqnarray}
p{}_M{}^n=Z_M{}^aJ_{\langle ab\rangle}Z^{bN}+Z_M{}^{\bar a}J_{\langle {\bar
a}{\bar b}\rangle}Z^{{\bar b}N}
~~&&~~
q{}_M{}^n=Z_M{}^{\bar a}J_{{\bar a}b}Z^{bN}+Z_M{}^{a}J_{{a}{\bar b}}Z^{{\bar
b}N}\nonumber\\
q'{}_M{}^n=Z_M{}^{b}J_{{\bar a}b}Z^{{\bar a}N}+Z_M{}^{\bar a}J_{ b{\bar
a}}Z^{bN}\nonumber
\end{eqnarray}
where $M$ and $N$ are $PSL(4|4)$ indices, $a$ and ${\bar a}$ are $Sp(4)$
indices and $\langle \rangle$ denotes the
antisymmetric and traceless part.
}
\begin{eqnarray} d {*p} &=&  p \wedge *q + *q \wedge p + {\textstyle
\frac{1}{2}}(q
\wedge q' + q' \wedge q)
\ ,\nonumber\\ 0 &=& p \wedge (*q - {q'}) +  (*q - {q'})
\wedge p\ ,\nonumber\\ 0 &=& p \wedge (q - *{q'}) +  (q - *{q'}) \wedge
p\ . \label{eom}
\end{eqnarray} Notice that
\begin{equation} d {*(p +{\textstyle \frac{1}{2}} *q')} = 0\ .
\label{curcon}
\end{equation} The conserved current $p +{\textstyle
\frac{1}{2}} *q'$ is the Noether current of the global left
multiplication symmetry.  The conservation equation~(\ref{curcon}) is
actually the complete equation of motion.   By converting to an equation
for $d {*(P +{\textstyle \frac{1}{2}}  {*Q'})}$ and decomposing the Lie
algebra one obtains all of eqs.~(\ref{eom}).

We now construct candidate connections.  We do not have equations for
$d{*q}$ and $d{*q'}$.  This is because the local $\kappa$ symmetry is
not yet fixed, so the equations of motion do not determine the full time
evolution.  In specific gauges one obtains equations for $d{*q}$ and
$d{*q'}$, but it turns out that we can construct the connection without
them.  Thus define
\begin{equation} a = \alpha p + \beta {*p} + \gamma q +
\delta q'\ .
\end{equation} Then, noting again the identities $** = +1$ and $*k
\wedge l + k \wedge *l = 0$, one finds
\begin{eqnarray}
da + a \wedge a &=& c_1 p \wedge p + c_2 (p \wedge q +
q \wedge p) + c_3  (p \wedge {q'} + {q'}
\wedge p) \nonumber\\
&&\qquad\qquad\qquad + c_4 q \wedge q + c_5 q'
\wedge q' + c_6(q
\wedge q' + q' \wedge q)\ ,
\end{eqnarray} where
\begin{eqnarray}
c_1 &=& -2\alpha + \alpha^2 - \beta^2 \ ,\nonumber\\
c_2 &=& -\alpha + \alpha\gamma - \beta\delta
\ ,\nonumber\\
c_3 &=& \beta - 2\delta + \alpha\delta -
\beta\gamma\ ,\nonumber\\
c_4 &=& {\textstyle
\frac{1}{2}}\alpha - 2\gamma + \gamma^2 \ ,\nonumber\\
c_5 &=&
{\textstyle \frac{1}{2}}\alpha + \delta^2 \ ,\nonumber\\
c_6 &=&
{\textstyle \frac{1}{2}} \beta  -\delta + \gamma\delta\ .
\end{eqnarray}

The vanishing of the $c_i$ gives six equations for four unknowns, but
there is a large degree of redundancy, and remarkably there are again
two one-parameter families of flat connections.  One can use the
vanishing of $c_5$ and
$c_6$ to solve for $\alpha$ and $\beta$, and the remaining
$c_i$ then all vanish if $\delta^2 = \gamma^2 - 2\gamma$. Thus the
connection $a$ is flat for
\begin{eqnarray}
\alpha &=& -2 \sinh^2 \lambda\ ,\nonumber\\
\beta &=& \mp 2 \sinh\lambda \cosh\lambda\ ,\\
\gamma &=& 1 \pm \cosh \lambda\ , \nonumber\\
\delta &=& \sinh \lambda \ .
\end{eqnarray} We do not have any deep understanding of why these flat
connections exist.  We have found $\kappa$ gauges in which there are two
fewer equations, but this still requires one redundancy.  There is some
connection between the $\kappa$ invariance and the nonlocal charges:
   if we rescale the Wess-Zumino term by $\sigma$ then both the
$\kappa$ invariance and the higher symmetries are broken (except for
$\sigma = -1$, which is just the world-sheet parity transform).

Notice that if we ignore fermions, the bosonic terms in
$p$ and $*p$ reproduce the currents for the bosonic coset (though only
with the lower sign for the latter).  This shows that the currents for
different $\lambda$ are independent --- they are not related to one
another by
$\kappa$ transformations.  Incidentally, the charges cannot be strictly
$\kappa$ invariant, because the generator of $\kappa$
transformations do not even commute with the global symmetries. It is
plausible to conjecture that all the commutators vanish weakly  --- that
is, they are themselves $\kappa$ transformations.

\sect{Local currents}

Consider again the bosonic examples, the group manifold or the coset,
where in each case there is a current (either $j$ or $2k$) which is both
flat and conserved.  In world-sheet light-cone components,
\begin{equation}
\partial_- j_+ = -\frac{1}{2} [j_-, j_+]\ .  \label{pmjp}
\end{equation} It follows immediately that~\cite{local,EHMM}
\begin{equation}
\partial_- {\rm Tr}(j_+^n) = 0\ .
\end{equation} Thus, even though the $G\times G$ currents themselves are
not chiral, there are higher spin chiral currents.  The traces are not
independent, so the number of higher spin currents is finite.  Similarly
there is a set of left-moving currents.

In the bosonic models, the classical scale invariance is broken by
quantum effects.  Similarly these chiral currents are anomalous, but
under appropriate conditions there will be a conserved though nonchiral
higher spin current~\cite{local}.  In the conformally invariant
supergroup models it has been argued that the chiral currents are
nonanomalous~\cite{BZW}.

In the $AdS_5 \times S^5$ case there is no current that is flat and
conserved, but the chiral currents can be constructed under weaker
conditions.  It suffices that
\begin{equation}
\partial_- j_+ = \sum_I [a^I, b^I]\ ,\quad [j_+,
a^I] = 0\ , \label{gencon}
\end{equation}
so that
\begin{equation}
\partial_- {\rm Str}(j_+^n)
= n  \sum_I {\rm Str}(j_+^{n-1} [a^I, b^I])
= n  \sum_I {\rm Str}([j_+^{n-1}, a^I], b^I) = 0\ .
\end{equation}

Let us try to proceed without fixing the $\kappa$ gauge.  Since we do
not have equations for
$d{*q}$ and $d{*q'}$, the only candidate for $j_+$ is
$p_+$.  Noting that $(*\omega)_\pm = \pm \omega_\pm$ for any one-form,
we find that
\begin{equation}
\partial_- p_+ = [p_+, p_- + q_-] - \frac{1}{4}[q_{1-}, q_{1+}]\ ,
\end{equation}
while the fermionic equations of motion become
\begin{equation}
[p_+, q_{1-}] = [p_-,q_{2+}] = 0\ . \label{feom}
\end{equation}
Together these imply that $\partial_- p_+$ is of the
form~(\ref{gencon}), with $a^1 = p_+$ and $a^2 = q_{1-}$.
Unfortunately these currents are actually trivial.  Since ${\rm
Str}(p_+^n) = {\rm Str}(P_+^n)$, we only need traces of the broken
bosonic generators.  By $SO(4,1)\times SO(5)$ invariance, these can only
involve products of $P_+^a P_+^a$ and $P_+^{a'} P_+^{a'}$, where $a$ is
an
$SO(4,1)$ vector index and $a'$ is an $SO(5)$ vector index.  In fact one
finds that
\begin{equation}
{\rm Str}(P_+^{2k}) \propto (- P_+^a P_+^a)^k - (P_+^{a'} P_+^{a'})^k\ .
\label{pkpk}
\end{equation}
The first minus sign is from the Minkowski signature of $AdS_5$, and the
second is from the supertrace.  For $k=1$ this is the world-sheet
$T_{++}$, which vanishes by the metric equation of motion, while for all
$k>1$ it
is a multiple of $T_{++}$.  Thus these would-be chiral currents vanish.
Notice however that if we go to conformal gauge, where the vanishing of
$T_{++}$ is not imposed as an equation of motion, then the fact that the
current~(\ref{pkpk}) is chiral for all $k$ implies that
$\partial_- (P_+^a P_+^a) = \partial_- (P_+^{a'} P_+^{a'}) = 0$
separately.

Of course, it could well be that there are currents that are chiral only
in certain gauges.  In the flat-spacetime case, for example, only after
fixing gauge are the world-sheet fields $\partial X$ and $\theta$ chiral.
A natural gauge for us is
\begin{equation}
q_{1-} = q_{2+} = 0\ .
\end{equation}
It should be possible to reach such a gauge, at least at the level of
the classical solutions: the equations of motion~(\ref{feom}) already
imply half of this, and there is enough gauge freedom to impose the rest.
In the flat-spacetime theory, the analogous gauge makes the world-sheet
fermions free.  Unfortunately, while the field and Maurer-Cartan
equations simplify substantially in this gauge, there are no additional
currents having the property~(\ref{gencon}).

\sect{Discussion}

The obvious next question is the use of these charges.  The
classic application is in theories with a mass gap, with the spatial
coordinate unbounded.  The relevant observable is then the $S$-matrix.
The higher spin local conservation laws~\cite{local} imply the
absence of particle production and the factorization of the $n$-particle
$S$-matrix in terms of two-particle $S$-matrices, allowing the full
$S$-matrix to be deduced (ref.~\cite{zz} and references therein).
These same constraints can be derived from the nonlocal
charges~\cite{luscher}.  The argument appears to be less
straightforward, but in at least some circumstances the local charges
can be regarded as limits of the nonlocal charges~\cite{brev}, so the
latter are sufficient.

These nonlocal charges also form the starting
point for the classical and quantum inverse scattering methods, which are
related in turn to the Bethe Ansatz (ref.~\cite{invrev} lists some
reviews).  We note that the existence of these charges is a nontrivial
fact: they do not exist in all cases (e.g. the non-symmetric-space
cosets), and in such cases when there is a mass gap there is presumably
particle production in scattering.  It is less clear how conformal
theories with the charges are distinguished.

We are interested in conformally invariant theories, on a bounded space
(the open or closed string).  In this case the simplest and most obvious
observable is the partition function.  This encodes the set of operator
dimensions (in spacetime conformal theories such as ${\cal N} =4$
Yang-Mills) or the meson and glueball masses (in confining theories).
In fact, this has been found for certain conformally invariant models
based on supergroups~\cite{readsal}.  This was done not by direct use of
nonlocal charges, but by related methods involving integrable lattice
models.  Thus, our result on the existence of these charges in the
$AdS_5 \times S^5$ theory should be taken as motivation to apply the
full set of methods of integrable field theory to this system.

For the nonlocal charges, the next step is to determine their classical
Poisson bracket algebra, and to extend this to the
quantum theory.  This requires us to deal with the $\kappa$ gauge
invariance (which we were largely able to sidestep), perhaps along the
lines of ref.~\cite{berk}.  The conformally invariant supergroup and
supermanifold models~\cite{sethi,BZW,BVW} provide interesting warmup
problems, without the complication of $\kappa$ symmetry.
It is also interesting to ask what is dual on the gauge side to the
symmetries that we have found.  Finally, the extension to less symmetric
and more QCD-like theories is challenging.

\subsection*{Acknowledgments}

We would like to thank Oliver DeWolfe, Andrei Mikhailov, and Anastasia
Volovich for helpful discussions.  This work was supported by
NSF grants PHY99-07949, PHY00-98395, PHY00-99590, and
PHY01-40151 and
DOE grant DOE91ER40618.

\end{document}